\documentclass[aps,prb,twocolumn,showpacs,superscriptaddress,nofootinbib]{revtex4-2}
\usepackage{graphicx,epsfig}
\usepackage{times,bbm}
\usepackage{graphics,dcolumn,bm,float}
\usepackage{amssymb,amsmath,rotate,color,amsfonts}
\usepackage[title,titletoc,toc]{appendix}
\usepackage{mathtools}
\usepackage{booktabs}
\usepackage{tcolorbox}
\usepackage[pagebackref=false,colorlinks,linkcolor=magenta,citecolor=blue,urlcolor=magenta]{hyperref}
\usepackage{wrapfig}
\usepackage{lipsum}
\usepackage{mwe}
\usepackage[mathlines]{lineno}
\usepackage{hyperref}
\usepackage{breqn}
\usepackage{bbold}
\usepackage{pgfplots}
\usepackage{float}
\usepackage{tkz-euclide}
\usepackage{braket}
\usepackage{physics}
\usepackage[caption=false]{subfig}
\usepackage[export]{adjustbox}
\usepackage{tikz}
\usetikzlibrary{through,calc}
\usetikzlibrary{positioning}

\newcommand{\bea}{\begin{eqnarray}}
\newcommand{\eea}{\end{eqnarray}}

\newcommand{\bb}[1]{\boldsymbol{#1}}
\newcommand{\nn}{\nonumber}

\begin{document}

\preprint{}

\title{Interplay of Tilt and Axion Fields in Topological Superconductors:\\ Anisotropy in the Meissner Effect}

\author{Mojtaba Hassani}
\author{Zahra Faraei}
\email{z.faraei@iasbs.ac.ir}
\affiliation{Department of Physics$,$ Institute for Advanced Studies in Basic Sciences (IASBS)$,$ Zanjan 45137-667311$,$ Iran}

\begin{abstract}
Topological superconductors host gapless surface states that fundamentally alter their electromagnetic response through the axion field term $\theta \bb E\cdot\bb B$, arising from the topological magnetoelectric effect. In this work, we investigate the electromagnetic properties of a three-dimensional topological Weyl superconductor by leveraging its theoretical mapping to a four-dimensional topological insulator with s-wave superconducting boundaries. By incorporating the tilt of Weyl cones into this model, we demonstrate that the tilt vector $\bb \zeta$ anisotropically modifies the axion field profile near the surface, leading to a tilt-enhanced Meissner effect and anomalous magnetic penetration depths. We show that the magnetic field component perpendicular to the tilt direction exhibits a non-exponential, hypergeometric decay dictated by the interplay between the axion term and $\bb\zeta$, while the parallel component remains largely tilt-insensitive—a hallmark of axion-mediated anisotropy absent in trivial superconductors. Remarkably, all tilt-dependent electromagnetic responses follow a universal scaling law, revealing a fundamental symmetry in the system's behavior. Furthermore, we predict a tilt-dependent planar Hall current at the surface, directly tied to the topological surface states.  
\end{abstract}

\date{\today}

\maketitle

\section{Introduction}
\label{sec.I}
Topological superconductors are a class of materials that exhibit superconductivity while also possessing unique topological properties~\cite{Topo_Ins_Sup,TSC_review2017,TSC_review2022,TSC_Materials}. These materials are characterized by a full pairing gap in the bulk and gapless surface states consisting
of Majorana fermions. This makes them promising candidates for fault-tolerant quantum computing~\cite{Ando,Quantum_Computation,Majorana_surface_codes,Majorana-based_qubits}. A key feature of superconductors, including topological ones, is the Meissner effect. This effect involves the expulsion of magnetic fields from the interior of a superconductor below its critical temperature, highlighting their unique electromagnetic properties~\cite{MeissnerEffect_1,MeissnerEffect_2}. In topological superconductors, the Meissner effect not only confirms the superconducting state, but also sheds light on the interaction between superconductivity and topologically protected surface states\cite{nogueira,witten2016axion,Hammar,massive_photon_propagator}.

The electromagnetic response and dynamics of superconducting phase fluctuations in topological superconductors are well described by the axion field theory, reflecting the robust theoretical framework inherent to this field \cite{witten2016axion,Gu_Qi,Bernevig_Book}.
This theoretical framework reveals that three-dimensional topological superconductors (3D TSC) exhibit a topological magnetoelectric effect, which arises from a term in the electromagnetic response of the form $\sum_i C_{1i} ~ \theta_i ~ \epsilon^{\mu\nu\tau\sigma} F_{\mu\nu} F_{\tau\sigma}$. Here, $i$ indexes all Fermi surfaces with a Chern number ($C_{1i}$), and $\theta_i$ denotes the superconducting phase associated with each Fermi surface \cite{Qi_witten}.

In a topological Weyl superconductor (TWSC) characterized by two Weyl cones in the normal phase, the Chern numbers associated with these cones, denoted as $C_R$ and $C_L$ for the left and right chirality respectively, are assigned values of $-1$ and $+1$. Consequently, the axion term is expressed as $(\theta_L - \theta_R) \epsilon^{\mu\nu\tau\sigma} F_{\mu\nu} F_{\sigma\tau}$, which is proportional to the phase difference between the order parameters of opposite chiralities, denoted as $\theta = \theta_L - \theta_R$. This term, which mirrors the coupling between axions and an Abelian gauge field~\cite{nenno}, underpins the designation of this theoretical framework as axion field theory. The effective action, including the axion term, is applicable to both topological and trivial superconductors, depending on the ground state values of $\theta_L$ and $\theta_R$. For a topological superconductor, $\theta_L-\theta_R = \pi$, while for a trivial superconductor, $\theta_L -\theta_R = 0$. The ground state value of $\theta_L - \theta_R$ is governed by the Josephson coupling between the phases of the pairing order parameters of opposite chiralities, which plays a crucial role in determining whether a superconductor is topological or trivial \cite{nogueira}.

In this paper, we aim to incorporate the tilting of Weyl cones into the theoretical framework of a 3D TSC to observe variations in the electromagnetic responses of the surface states. To achieve this, we employ the mapping of a 3D TSC to a four-dimensional topological insulator (4D TI) with s-wave superconductors on the boundaries. This theoretical tool allows us to leverage the well-understood properties of TIs to explain and predict the behavior of TSCs\cite{Qi_witten,PRB78.195424,Gu_Qi,witten2016axion,PRX2.031008,Ryu_2010}. This approach provides a framework to understand the key role of superconducting phase fluctuations and the axion coupling term in the physics of TSCs.  The analogy uses the concept of an extra spatial dimension to link the topological properties of the two materials. While not a literal physical construction, it serves as a conceptual analogy that is useful for understanding the electromagnetic response and topological classification of TSCs. The details of this mapping and the effective Lagrangian are provided in Appendix A. We assume that the two Weyl superconductors on opposite sides of the 4D TI model exhibit moderate tilts. We will then investigate how this tilt affects the electromagnetic responses of the entire system.

In the study of Weyl materials, particular attention is given to those with Weyl cones that exhibit a tilt within the Brillouin zone, as documented in several studies~\cite{Tilted_Weyl,Tilt_Bernevig,TaIrTe,MoWTe,WTe}. The generation of this tilt is often linked to the atomic structure~\cite{Goerbig_tilt,tilt_atomic_structure,tilt_intrinsicStructure} and also the application of strain or pressure in Weyl semimetals\cite{tilt_strain,tilt_strain_torsion,Curved_spacetime_theory}. We will show that the tilt of the Weyl cone plays a significant role in dictating the profile of the axion field near the surface, which in turn affects the magnetic field penetration within TWSCs. Tilted Weyl semimetals have been predicted to exist in a variety of materials, and were predicted to show a spectrum of properties in the normal phase(non-superconducting phase) of the Weyl semimetals, including but not limited to anisotropy in the planar Hall effect, electrical conductivity, magnetotransport, thermal and optical properties, and the Nernst effect~\cite{tilt_Hall,tilt_magnetic2,anisotropy_conductivity,tilt_magnetic,tilt_magnetic2,tilt_magnetoptical,Theo_TWSM,tilt_nernst,tilt_optical,tilt_optical2,tilt_optical3,tilt_nernst2}.  However, the focal point of our inquiry is the influence of the Weyl cone's tilt in the superconducting phase. We delve into the tilt impact on the electromagnetic response of a TWSC—a phenomenon that remains undetected in trivial superconductivity but becomes apparent through the intermediary of the axion field (evidenced by gapless boundary states on the surface). Thus, we propose that this anisotropy is an axion-mediated phenomenon. 

To confirm our proposition, we have incorporated the Weyl cone tilt as a metric alteration within the Lagrangian framework of the TWSC~\cite{Tilt_Curved-Space,SaharPolariton}. We have further concentrated on the consequences of tilting for the magnetic field penetration and the planar Hall effect on the surface of the topological Weyl  superconductor. The scope of our analysis is confined to moderate tilts ($\zeta < 1$), which are characterized by the preservation of a point-like Fermi surface.

This paper is structured as follows: Sec.~\ref{sec.II} provides an overview of the axion field in topological superconductors, detailing its computation for systems with a tilted Weyl structure.  Sec.~\ref{sec.III}  presents the spatial profile of the axion field, resolving its dependence on the Weyl cone tilt vector. Secs. \ref{sec.IV} and \ref{sec.V} outline the modified magnetic penetration patterns, distinguishing between perpendicular and parallel field components relative to the tilt axis. In Sec. \ref{sec.VI}, we explore the scale invariance behavior of the magnetic field with the tilt vector. Sec.\ref{sec.VII} examines the emergent anisotropic planar Hall effect and its associated surface currents. The tilt-dependent modulation of the superconducting order parameter is systematically explored in Sec.\ref{sec.VIII}. Finally, Sec.\ref{sec.IX} summarizes our principal findings.

\section{Model and Formulas }
\label{sec.II}
In the vicinity of a Weyl point, the Hamiltonian characterizing a tilted Weyl semimetal is given by
\bea
H_\chi=\chi \bb\sigma \cdot \bb p + \sigma_0~{\bb\zeta}_\chi \cdot \bb p,
\eea
with $(\sigma_0, \vec{\sigma})$ denoting the $2\times 2$ Pauli matrices in the space of the two crossing bands. The chirality is denoted by $\chi=\pm$, and ${\bb\zeta}_\chi=(\zeta_x,\zeta_y)_\chi$ is the tilt vector, indexed by $\chi$ to account for the scenarios where the tilts are “towards each other” and “away from each other.”

The spontaneously symmetry breaking  in quantum field theory can be demonstrated through the
theory of a complex scalar field $\phi$ with the Lagrangian,
\bea
\label{L}
\mathcal{L}=(\partial_\mu \phi)^\dagger (\partial^\mu \phi) + m^2 \phi^\dagger \phi - (\frac{\lambda}{2}) (\phi^\dagger \phi)^2 .
\eea
In this context, both $m^2$ and $\lambda$ in the condensation part are positive and the stable minima are located at $\phi^\dagger \phi= m^2/\lambda$. The kinetic term, $(\partial_\mu \phi)^\dagger (\partial^\mu \phi)$ is defined with  $\partial^\mu \phi:=g^{\mu\nu} \partial_\nu \phi$ where $g^{\mu\nu}$ is the inverse of the $4\times 4$ metric tensor of the space-time coordinates. The tilting of Weyl cones introduces a tilt-dependent inverse metric represented by the matrix~\cite{SaharPolariton}, 
\bea
g^{\mu\nu}=
\begin{pmatrix}
    1 & \zeta_x & \zeta_y & \zeta_z 
    \\
    \zeta_x   &  -1+\zeta_x^2     &  \zeta_x \zeta_y     & \zeta_x \zeta_z 
    \\
    \zeta_y   &  \zeta_y \zeta_x     &  -1+\zeta_y^2     & \zeta_y \zeta_z 
    \\
    \zeta_z   &  \zeta_z \zeta_x     &  \zeta_z \zeta_y     & -1+\zeta_z^2
\end{pmatrix},
\label{inversemetric}
\eea
which in $\bb\zeta=0$ becomes the Minkowski metric $\eta^{\mu\nu}=\eta_{\mu \nu}=\text{diag}[1,-1,-1,-1]$. 

Type-I Weyl semimetals are characterized by moderate tilts, $|\zeta| < 1$, while type-II ones exhibit over-tilt values, $|\zeta| > 1$. The components $\zeta_x$, $\zeta_y$, and $\zeta_z$ represent the tilt projections in the $x$, $y$ and $z$ directions, respectively, with the overall tilt magnitude given by $\zeta=\sqrt{\zeta_x^2+\zeta_y^2+\zeta_z^2}$.

In a TWSC hosting at least two distinct Fermi surfaces, the system is characterized by two superconducting order parameters, $\Delta_+$ and $\Delta_-$,
associated with the respective Fermi surfaces. These order parameters exhibit a relative phase difference of $\pi$, reflecting a nontrivial interplay between the Cooper pairing channels. For instance, a spin-singlet pairing potential of the form $\Delta_\chi= i\Delta_{0\chi} e^{i\theta_\chi} \sigma_y$ ($\chi=\pm$) with $\theta_+-\theta_-=\pi$, gives rise to a parity-odd pseudo-scalar Weyl superconductor. Here, the parity transformation ($\chi\rightarrow -\chi$) flips the sign of the scalar $\Delta_\chi$, enforcing $\Delta_{0+}=\Delta_{0-}$ and ensuring the system’s topological character.

The total free energy (here, the free energy corresponds to the Lagrangian density in the field-theoretic formulation, as we account for both static thermodynamic potentials and dynamic electromagnetic fields) of the system comprises three contributions: (i) the intrinsic free energy of each superconducting condensate, (ii) the Josephson coupling energy between the two order parameters, and (iii) the electromagnetic field energy in the presence of external fields. Crucially, a topological term emerges in the Lagrangian, proportional to the phase difference $\theta=\theta_+-\theta_-$ between the two order parameters. This term couples to the electromagnetic field tensor $F_{\mu\nu}$ in a manner analogous to the axion electrodynamics term $\theta F_{\mu\nu} \tilde{F}_{\mu\nu}$, where $\tilde{F}_{\mu\nu}$ is the dual field tensor. Unlike conventional superconductors, where such a term vanishes identically, its presence here is a hallmark of the topological nature of the superconductor.

The inclusion of this topological term modifies the electromagnetic response of the system, leading to anomalous behavior in the Meissner effect at the boundaries of the TWSC~\cite{nogueira}. Specifically, the screening of magnetic fields deviates from the standard exponential decay observed in trivial superconductors, reflecting the interplay between topology and symmetry breaking.

To rigorously derive the effective Lagrangian governing these phenomena, we employ a dimensional mapping between the 3D TWSC and a 4D TI. This correspondence is rooted in the shared symmetry classification of the two systems—both belong to the same homotopy group in the Altland-Zirnbauer scheme—and the equivalence of their surface state structures under dimensional reduction. For instance, the Majorana boundary modes of the 3D TWSC correspond to the Dirac fermion states of the 4D TI, while the bulk topological invariants (e.g., the 3D Chern-Simons coefficient and the 4D second Chern number) are related through compactification of the fourth spatial dimension. The explicit construction of the effective Hamiltonian and its mapping to the 4D TI framework are detailed in Appendix A.

The mapping of a 3+1D TSC to a 4+1D TI is achieved by coupling the 4+1D TI’s surface states to s-wave superconductors with phases $\theta_-$ and $\theta_+$ on opposite boundaries. The phase difference $\theta=\theta_+-\theta_-$ maps to a vector potential $A_4$ in the extra spatial dimension, and a gauge transformation incorporates $\theta$ into the electromagnetic response. The resulting effective action includes a topological axion term $\theta (\bb E \cdot \bb B)$, derived from the second Chern number $C_2$ of the 4+1D TI, which governs the magnetoelectric response of the TSC. This framework connects the topological invariants of the 4+1D TI (e.g., $C_2=1$) to those of the 3+1D TSC, with $\theta=0$ or $\theta=\pi$ determining the TSC’s topological phase. The Lagrangian combines the axion term, Maxwell term, Higgs term, and Josephson coupling, providing a unified description of the TSC’s electromagnetic and topological properties.

By leveraging this mapping, we arrive at the following effective Lagrangian for the TWSC\cite{Qi_witten, nogueira}:
\bea
\nn
\mathcal{L} =
&&\sum_{\chi=\pm} \bigg[ 
(\partial_\mu\theta_\chi - q A_\mu)^\dagger ( \partial^\mu\theta_\chi - q A^\mu ) |\Delta_\chi|^2 
\\
\nn
&& + m^2 |\Delta_\chi|^2  - \frac{\lambda}{2} |\Delta_\chi|^4 \bigg] \\
&&+~ \gamma~(\Delta_+^* \Delta_- + \Delta_-^* \Delta_+)\\
\nn
&&- \frac{1}{4} F_{\mu\nu} F^{\mu\nu} + \frac{e^2(\theta_+ -\theta_-)}{64\pi^2} ~ \epsilon^{\mu\nu\alpha\sigma} F_{\mu\nu} F_{\alpha\sigma}.
\label{main_L}
\eea
The first two terms in the Lagrangian describe the superconducting sector, where the coupling to the electromagnetic sector is implemented via minimal coupling with charge $q=2e$. Throughout this work, we adopt natural units ($\hbar=c=1$), reintroducing these constants when necessary.

The Josephson term, parameterized by its coupling strength $\gamma$, emerges due to the coexistence of two order parameters with distinct phases, $\theta_+$ and $\theta_-$. To construct the Lagrangian describing the superconducting components, we employ the two-component Ginzburg-Landau (TCGL) model~\cite{babaev2012}, which is consistent with the system's single $U(1)$ symmetry. Furthermore, we assume the coefficients associated with the two order parameter components to be equal, a simplification justified by the equal electron density for both chiralities in the system. This assumption ensures a balanced treatment of the superconducting order parameters, reflecting the symmetric nature of the underlying electronic structure.We further assume a constant coupling amplitude, $\Delta_+=\Delta_-=\Delta_0$, for the superconducting order parameters. This simplification is justified as our primary focus is on the electromagnetic response, which is predominantly coupled to the phase dynamics of the order parameter. Consequently, we have neglected contributions arising from fluctuations in the amplitude of the order parameter, as they do not play a significant role in the axion-mediated electromagnetic response under investigation.

The term $-1/4 F_{\mu\nu} F^{\mu\nu}$ represents the electromagnetic free energy, while the coefficient $e^2 / 16\pi^2$ in the last term quantifies the strength of the topological coupling. This Lagrangian provides a unified framework for understanding the anomalous electromagnetic response of the TWSC, including its modified Meissner physics.  Further derivations leading to Eq. \eqref{main_L} are provided in Appendix A. It is important to note that the topological term is physically meaningful only in the superconducting state (where $|\Delta| \ne 0$). In the limit $|\Delta| \rightarrow 0$, the phases $\theta_+$ and $\theta_-$ become undefined, and the topological contribution vanishes, as the system reverts to the normal state.

The tilt effect is hidden within the contravariant form of the electromagnetic field tensor, $F^{\mu\nu}=g^{\mu\alpha} g^{\nu\sigma} F_{\alpha\sigma}$, the contravariant gradient four-vector $\partial^\mu=g^{\mu\nu}\partial_\nu$, and the vector potential $A^{\mu}=g^{\mu\nu}A_{\nu}$, all defined  using the inverse metric (\ref{inversemetric}).  For now, we neglect the tilt dependence of $\Delta_0$, though we revisit this in Sec. \ref{sec.VII} and demonstrate that its impact remains minor within the broader scope of our analysis.

By performing a gauge transformation $A_\mu \rightarrow A_\mu+ (1/2q) \partial_\mu (\theta_+ + \theta_-)$, and changing variables from $(\theta_+,\theta_-)$ to $(\theta,\varphi)$ where $\theta=\theta_+ -\theta_-$ and $\varphi=\theta_+ +\theta_-$, we arrive at a more convenient form for the Lagrangian:
\bea
\nn
\mathcal{L} &=&
\frac{\Delta_0^2}{4} 
(  \partial_\mu\theta  \partial^\mu\theta + 4~ q^2 A_\mu A^\mu )
\\
&+& 2~ \gamma ~ \Delta_0^2 \cos{\theta} 
+ m^2 \Delta_0^2
- \frac{\lambda}{2} \Delta_0^4 \\
\nn
&-& \frac{1}{4} F_{\mu\nu} F^{\mu\nu} + \frac{e^2\theta}{64\pi^2} ~ \epsilon^{\mu\nu\alpha\sigma} F_{\mu\nu} F_{\alpha\sigma}.
\label{L_con}
\eea
This transformation clearly shows that while one of the degrees of freedom is gauged away with the electromagnetic field, the other remains within the theory.

 We derive the equation of motion for this remaining degree of freedom, termed the 'axion field', due to its coupling with the electromagnetic field. This coupling mirrors the interaction between the axion field and an Abelian gauge field in quantum field theory~\cite{Wilczek_axion}.  Notably, we consider static phase fields, meaning $\partial_t \theta_\pm=0$ and  $\partial^j=(-1+\zeta_j^2)\partial_j + \sum_{\ell\ne j}\zeta_j\zeta_\ell \partial_\ell$ with $j,\ell=1,2,3$. Consequently, the opposite sign of the $\bb\zeta$ vector for different chiralities is irrelevant.

The Euler-Lagrange equations for the field $\theta$ and the gauge field are given by:
\begin{subequations}
\bea
\label{teta_dif}
&&\nabla^2\theta + \frac{1}{\lambda_\theta^2} \sin{\theta} = 
 \frac{e^2}{4\pi^2 \Delta_0^2} \bb E \cdot \bb B 
 \\
\label{F_dif}
&& \partial_\mu F^{\mu\nu} -
\frac{e^2}{8\pi^2} \epsilon^{\mu\nu\alpha\sigma} F_{\alpha\sigma} \partial_\mu \theta =j^\nu .
\label{curl-B}
\eea
\end{subequations}
Here, $\lambda_\theta=1/\sqrt{4\gamma}$ denotes the penetration depth of the axion field, while $j^\nu=q\Delta_0^2 (\partial^\nu\varphi-2q A^\nu)$ represents the current density. Under our gauge transformation, this reduces to $j^\nu=-(1/\lambda_B^2) A^\nu$, where $\lambda_B=1/\sqrt{2}q\Delta_0$. As we will later demonstrate, $\lambda_B$ corresponds to the magnetic penetration depth in the absence of both tilt and the axion field. 

Notably, the tilt modifies the dynamics of the $\theta$-term via the $\nabla^2=\partial_\mu\partial^\mu$ operator, thereby altering the electromagnetic field response - a distinctive feature of topological superconductors that is absent in their topologically trivial counterparts. Furthermore, the tilt parameter introduces additional modifications to the electromagnetic field through the contravariant (upper-index) terms in equation \eqref{F_dif}. These latter modifications represent generic effects that occur in both topological and non-topological superconductors.

To elucidate the tilt dependence, we explicitly rewrite Eq.\eqref{F_dif} for $\nu=1,2,3$, highlighting the contributions from the tilt parameter: 
\bea
\nn
&-&(1-\zeta^2)\bb\nabla\times\bb B 
-\bb\nabla\times[\bb\zeta(\bb\zeta\cdot\bb B)] 
-\bb\nabla\times(\bb\zeta\times\bb E) 
\\
\nn
&-&
\frac{1}{\lambda_B^2} [\bb A-\bb\zeta(\bb\zeta\cdot \bb A)]+\partial_t\bb E + \bb\zeta\times\partial_t \bb B\\
&=&
- \frac{e^2}{\pi} (\bb\nabla\theta \times \bb E).
\label{B.eqn}
\eea
To develop deeper physical intuition, we solve these equations for a simplified geometry that explicitly reveals the anisotropic nature of the electromagnetic response. 

We consider a half-infinite topological superconductor in the region $z>0$, subjected to parallel surface magnetic ($\bb B_0$) and electric ($\bb E_0$) fields  both lying in the $xy$-plane, along with a tilt vector $\bb\zeta=(\zeta_x,\zeta_y)$. In this geometry, by applying the curl operator and time derivatives to Eq. \eqref{B.eqn}, we obtain:
\begin{subequations}
\bea
&& \nabla^2 \bb E - \frac{1}{\lambda_B^2} \bb E = 0,
\label{E-u}
\\
\nn
&& (1-\zeta^2)\nabla^2 \bb B +\bb\zeta(\bb\zeta\cdot\nabla^2\bb B) 
-\frac{1}{\lambda_B^2} \bigg[ \bb B - \bb\zeta\times( \bb B \times\bb\zeta )\bigg] 
\\
\label{B-u}
&&= 
-\frac{e^2}{\pi} \bb\nabla\times 
(\bb\nabla\theta\times\bb E).
\eea
\end{subequations}

The full derivation is detailed in Appendix B. As indicated by Eq.(\ref{B-u}), the profile of the axion term near the surface significantly influences the magnetic field profile. We consider two components for $\bb B$ and $\bb E$. One parallel to the tilt vector and one perpendicular to it. The electric field is:
\bea
\bb E = \bb E_0  e^{-z/\lambda_B}
\label{E.eqn}
\eea
For the magnetic field:
\bea
\label{B-d-zxzy}
&&  \frac{d^2 B_\parallel}{dz^2} - \frac{1}{\lambda_B^2} B_\parallel =  
\frac{e^2}{\pi} \frac{d}{dz} (E_\parallel \frac{d\theta}{dz})\\
\nn
&& \frac{d^2 B_\perp}{dz^2} - \frac{1}{\lambda_B^2} B_\perp =\frac{e^2}{\pi(1-\zeta^2)} \frac{d}{dz} (E_\perp \frac{d\theta}{dz}).
\eea
As indicated by Eq.~(\ref{B-d-zxzy}), the profile of the axion term near the surface, significantly influences the perpendicular component of magnetic field profile.  Consequently, the tilt vector $\bb\zeta$  plays a crucial role in shaping the magnetic field profile and the Meissner effect in topological superconductors. In the absence of the axion term, as in trivial superconductors, Eq. (\ref{B-d-zxzy}) reduces to a homogeneous differential equation with the trivial solution $\bb B_0 e^{-z/\lambda_B}$, where $\bb B_0$ is the magnetic field at the surface. This suggests that $\bb\zeta$ does not affect the electromagnetic response in trivial superconductors lacking the axion term.

It is noticeable that in our model, the tilt vector has only $x$ and $y$ components. If $\zeta_z\ne 0$, the tilt would also affect the magnetic field inside the trivial superconductor as detailed in Appendix C. This effect is present in both trivial and topological superconductors. However, since our focus is on differences caused by surface states, we set $\zeta_z=0$.  In the following, we discuss the behavior of the magnetic field components in the presence of the axion term or ``axion-mediated" Meissner effect.

\section{Axion field penetration profile}
\label{sec.III}
The differential equation \eqref{teta_dif} approximates the equation of motion for a particle in a tilted washboard potential, represented as $-\zeta_\theta \theta+(1/\lambda_\theta^2)\cos\theta$, where $\zeta_\theta=e^2\bb E\cdot \bb B/4\pi^2 \Delta_0^2$ denotes the potential tilt. Thus, the tilt vector $\bb\zeta$ determines the potential tilt experienced by $\theta$. However, while the tilt of the Weyl cone remains constant, the washboard potential's tilt for $\theta$ in our model varies with $z$, the distance from the surface.

In this analogy, the solution of Eq.\eqref{teta_dif} is influenced by the parameter $\zeta_\theta$. For small values of $\zeta_\theta ~ (i.e., |\zeta_\theta|< 1/\lambda_\theta^2)$, the axion particle is trapped in a local minimum, resulting in $\theta$ taking a constant value of $n\pi$, where $n$ is an integer. The axion field may oscillate around this local minimum (the initial value of $\theta=\pi$ on the surface adds $\pi$ to all subsequent values of $\theta$).  

For larger values of $\zeta_\theta ~ (i.e., |\zeta_\theta|> 1/\lambda_\theta^2)$, the axion particle escapes these minima, moving freely down the potential slope. Deep within the superconductor, where $z\rightarrow\infty$, and the tilt of the washboard potential approaches zero, $\theta$ assumes a constant value of $2\pi n$ where $n$ is an integer (considering the initial conditions). This behavior aligns with our understanding of the axion field, which originates from the gapless boundary states and approaches zero (or equivalently $2\pi n$) deep within the topological superconductor.

\begin{figure}[t]
\includegraphics[width=5cm]{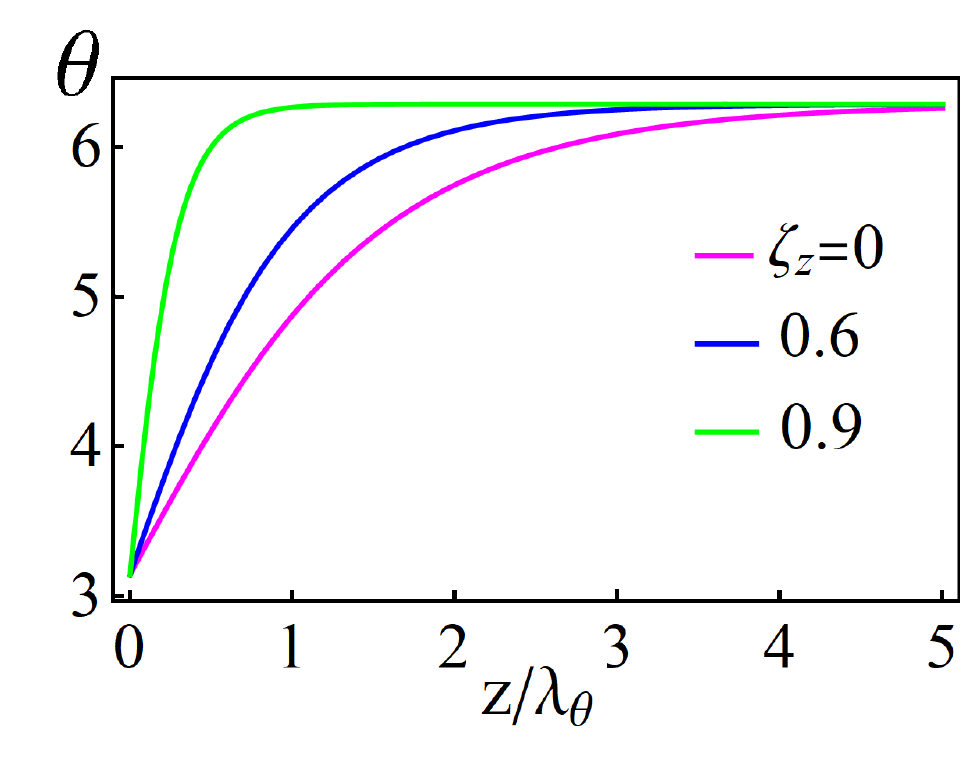}
\caption{Penetration profiles of topological surface states for varying tilt magnitudes $\zeta_z$. The axion field $\theta$ exhibits a $2\pi$-periodicity with $\theta=0$ and $\theta=2\pi$ both corresponding to trivial (non-topological) superconducting phases. }
\label{theta.fig}
\end{figure}

To streamline the analysis, as suggested in \cite{nogueira}, we omit the small non-homogeneous part of the differential equation for $\theta$ and consider the solution of the homogeneous equation adequate: $\theta=\pi+2\arcsin[\tanh(z/\lambda_\theta)]$. Fig \ref{theta.fig} illustrates the variation of $\theta$ with the distance from the surface inside the topological superconductor for different tilt magnitudes. Larger tilts narrow the region of influence of the surface states, resulting in a steeper $\theta$ curve. This steep gradient significantly impacts the magnetic field behavior in this region.

With the axion field profile determined, we now investigate the magnetic field behavior near the surface. We analyze two distinct cases: (1) the magnetic field component perpendicular to the tilt vector and (2) the parallel component. As evident from Eqs.\eqref{B-d-zxzy}, these components decouple from each other, allowing for separate analysis.

\section{perpendicular component of the magnetic field}
\label{sec.IV}
We first analyze the configuration where both surface-applied magnetic and electric fields are perpendicular to the tilt vector. In this case, the initial condition $E^\parallel_0 = 0$ enforces a vanishing parallel electric field component ($E_\parallel = 0$). Consequently, Eq. \eqref{E.eqn} reduces to $B_\parallel(z) = B^\parallel_0 e^{-z/\lambda_B}$ for the parallel magnetic field component. However, this component vanishes completely due to the boundary condition $B_0^\parallel=0$.

For the perpendicular component of $\bb B$, we solve the second equation of Eq.\eqref{B-d-zxzy} with $E_\perp = E_0 \exp(-z/\lambda_B)$. Eq.\eqref{B-d-zxzy} reveals that the tilt modifies the magnetic profile through the axion term. In trivial superconductors, where the axion term is absent, Eq.\eqref{B-d-zxzy} reduces to a homogeneous differential equation with the simple solution $\bb B_0 e^{-z/\lambda_B}$. However, in topological superconductors—where the axion field is present—the exact solution for $\bb B_\perp$ takes a more intricate form:
\bea
\label{analyticalBperp}
B_\perp &=&  B_0 ~ e^{-z/\lambda_B} \\
\nn
&+& \frac{e^2 E_0 ~ e^{-z/\lambda_B}}{\pi(1-\zeta^2)}
\bigg\{ \arcsin{[\tanh{(z/\lambda_\theta)}]} -
\\
\nn
&& \frac{2a}{a+2} e^{-z/\lambda_\theta} 
~_2F_1(1,\frac{a+1}{2a},\frac{3a+2}{2a},-e^{-2z/\lambda_\theta})
\bigg\},
\eea
with $a=\lambda_\theta/\lambda_B$  representing the ratio of the penetration depth of the axion field to the trivial penetration depth of the magnetic field. This solution reveals two physically distinct contributions to the magnetic response:
The first term ($B_0 ~ e^{-z/\lambda_B}$) represents the conventional Meissner effect, universal to all superconductors.  The remaining terms describe the axion-mediated magnetic field - a  signature of topological superconductivity absent in trivial systems. This contribution arises from the applied electric field and vanishes when $\bb E_0=0$, reflecting its origin in the fundamental axion-electrodynamic coupling $\theta (\bb E \cdot \bb B)$, which intrinsically links the axion field to the electromagnetic configuration.

\begin{figure}[t]
\includegraphics[width=6.5cm]{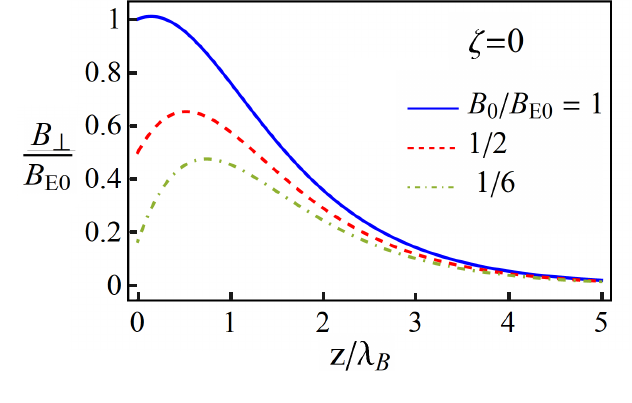}
\caption{Magnetic field profile perpendicular to the tilt direction as a function of $z$, the distance from the surface inside the superconductor, for varying initial values of $B_\perp$ at the surface. Here, $z$ is scaled with $\lambda_B$, the penetration depth of the magnetic field in trivial case, and $B_\perp$ is scaled with $B_{E_0}=e^2 E_0/\pi$, where $E_0$ is the applied electric field at the surface. The contribution of the axion-mediated electrically induced magnetic field increases with rising $E_0$. This enhancement transforms the magnetic field profile from an exponential decay function to a hypergeometric function with a broad peak near the surface.}
\label{B.fig}
\end{figure}

Even in the absence of tilting ($\zeta = 0$), a topological superconductor exhibits modified Meissner physics when $e^2E_0/\pi$ becomes comparable to $B_0$. Restoring the fundamental constants, this condition corresponds to $(24\pi^2\hbar c~ \epsilon_0~\mu_0/e^2) B_0/E_0 \lesssim 1$. As shown in Fig.~\ref{B.fig}, when this ratio approaches or drops below unity, the magnetic field penetration in the topological superconductor significantly deviates from conventional exponential screening. For typical experimental parameters—a magnetic field on the order of \text{mT} and an electric field of $\sim 10^5$ V/m—the ratio becomes approximately $1/10$, which is sufficient to observe the axion-mediated magnetic field penetration. Note that a single polarized photon beam with orthogonal $\mathbf{E}$ and $\mathbf{B}$ fields is insufficient; at least two orthogonal beams are required to generate parallel electric and magnetic field components.

This explanation provides an intuitive understanding of axion-induced photon interactions, as discussed in \cite{nogueira}, which are absent in non-topological superconductors. The Anderson-Higgs mechanism is usually interpreted as the creation of massive photons through the absorption of fluctuations in the phase of the order parameter. However, in topological superconductors, an additional phenomenon occurs. With two phases present, photons absorb the fluctuations of one phase, thereby acquiring mass, while the fluctuations of the other phase induce photon-photon interactions.

Fig.~\ref{B.fig} depicts the magnetic field profile near the surface for $\zeta=0$ but varying $B_0/E_0$ ratios. The dominance of the induced magnetic field, indicative of the axion field's effectiveness, is contingent upon this ratio.

\begin{figure}[t]
\includegraphics[width=8cm]{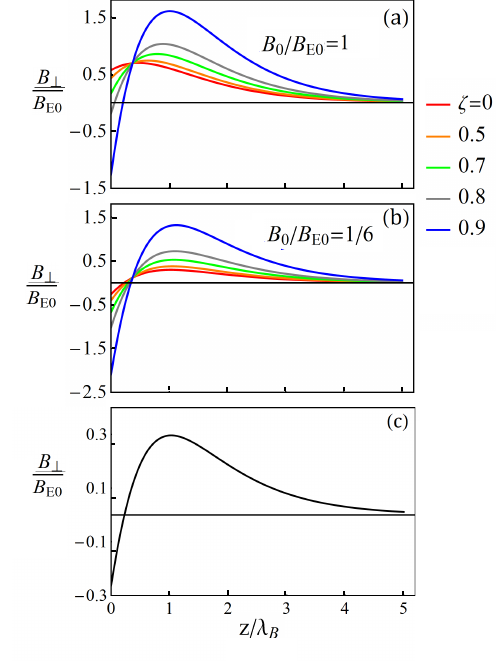}
\caption{Magnetic field profile perpendicular to the tilt direction as a function of $z/\lambda_B$ for varying tilt magnitudes. In (a) the initial surface value of $B_\perp$ is $B_0=B_{E0}$, while in (b) $B_0=(1/6) B_{E0}$. (c) All data from the panel (b), and additional data for a system with $B_0/B_{E_0}=1/6$, fall into a universal curve for $(1-\zeta^2)B_\perp/B_{E_0}$ upon the scaling $z\rightarrow \sqrt{1-\zeta^2}~ z$ and $\lambda_{B(\theta)}\rightarrow\sqrt{1-\zeta^2} ~ \lambda_{B(\theta)}$.}
\label{tilt.fig}
\end{figure}

\begin{figure}[t]
\includegraphics[width=7cm]{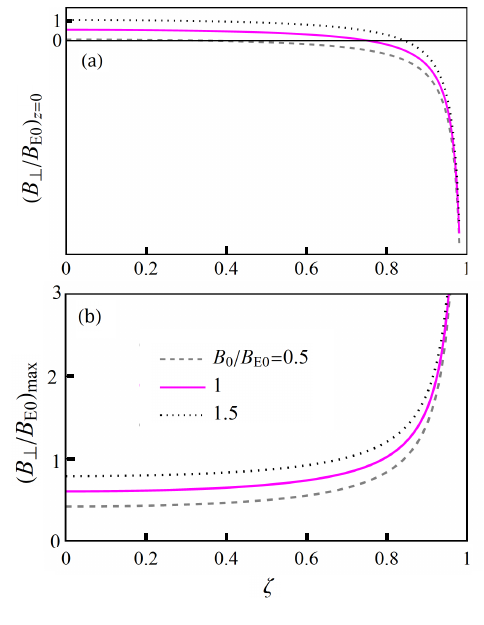}
\caption{Magnetic field inside the superconductor as a function of $\zeta$ at $z=0$ and $z=\lambda_B$, where $B_\perp$ reaches its maximum value. Various magnitudes of the applied magnetic field are considered. Increasing the applied magnetic field reduces the contribution of the surface states, while tilting enhances this contribution.}
\label{peak.fig}
\end{figure}

\begin{figure*}[t]
\includegraphics[width=18cm]{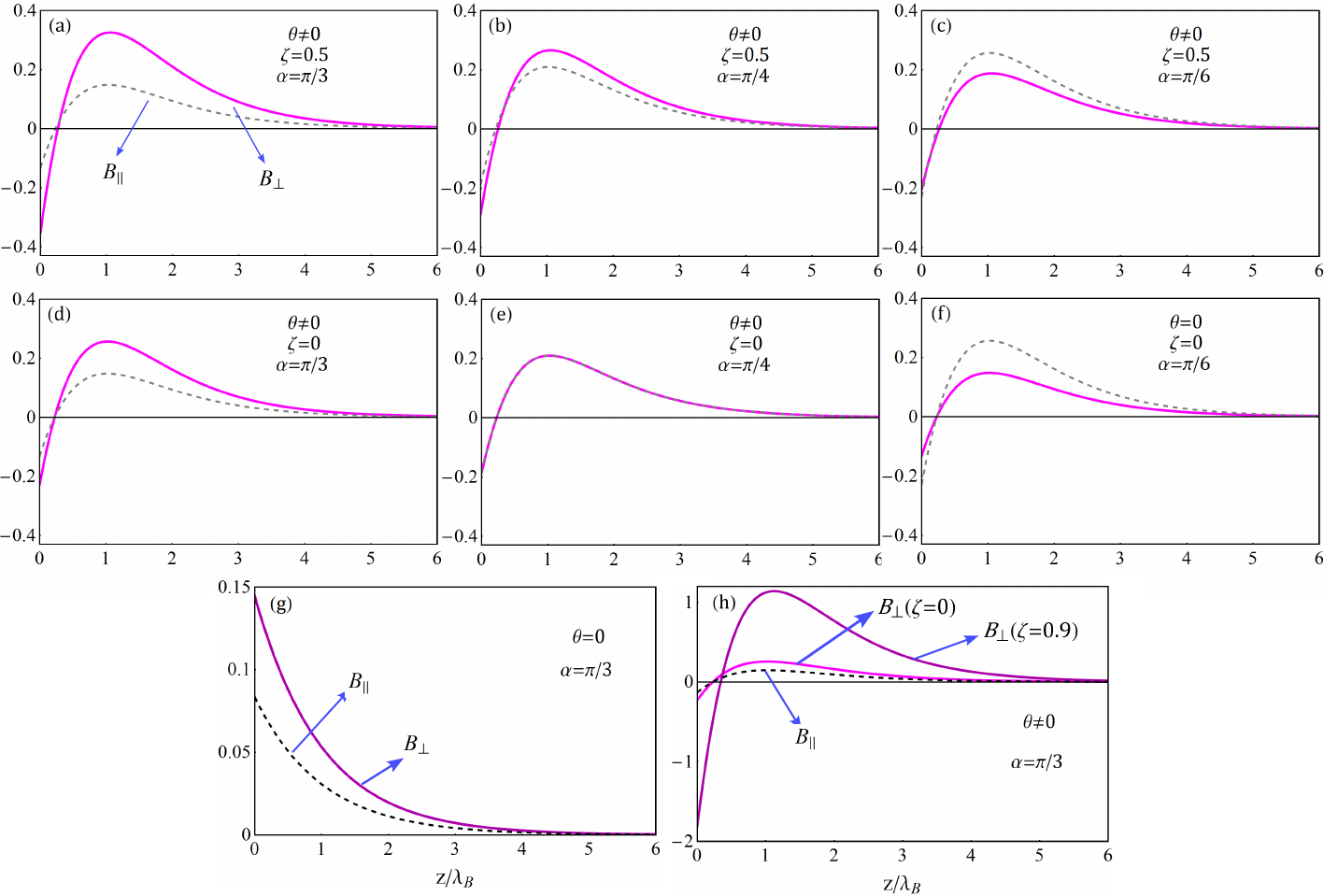}
\caption{Magnetic field behavior near the surface, parallel and perpendicular to the tilt direction. All vertical axes show $z/lambda_B$. Magenta solid curves represent $B_\perp$, while gray dashed curves denote $B_\parallel$. (a)–(c): $B_\perp$ and $B_\parallel$ in a TWS with fixed $\zeta=0.5$ and varying $\alpha$. (d)–(f): $B_\perp$ and $B_\parallel$ in a TWS with fixed $\zeta=0$ and varying $\alpha$. (g) Trivial superconductor case ($\theta=0$), where both $B_\perp$ and $B_\parallel$ exhibit identical exponential decay, independent of $\zeta$. (h) TWS case revealing anisotropic penetration: $B_\perp$ shows tilt-dependent modification, while $B_\parallel$ remains unaffected by $\zeta$.}
\label{anisotropy.fig}
\end{figure*}

The presence of tilt amplifies the electrically induced magnetic field, enabling it to manifest even at weaker external electric fields.  Fig.~\ref{tilt.fig} (a) and (b) display the magnetic field penetration profiles for a fixed $B_0/E_0$ ratio but varying tilt strengths. These results demonstrate that the tilt parameter significantly enhances the effect of the axion field. Additionally, the figures reveal that larger tilt values lead to an increased magnetic field penetration depth. As the tilt grows stronger, the field penetrates deeper into the superconductor, tending to zero at greater distances from the surface.

The ratio $B_0/E_0$ significantly influences the behavior of the magnetic field in a topological superconductor. However, at a certain threshold, it is the magnitude of the tilt that determines where the exponential behavior is dominated by the induced magnetic field from the external electric field. As the tilt increases, the exponential behavior diminishes progressively. It appears that a larger tilt enhances the influence of surface states, allowing the electromagnetic field to surpass the external magnetic field in dominance. Consequently, in addition to the $B_0/E_0$ ratio, the magnitude of the tilt emerges as a critical parameter that accentuates the presence of surface states, making them more pronounced.

To illustrate this issue more clearly, in Fig.~\ref{peak.fig}, we compare the magnetic field inside the superconductor relative to the tilt magnitude at two critical points for different $B_0/E_0$ values. 

The significance of $z=0$ lies in the fact that the external electric field induces a magnetic field in the opposite direction of the external magnetic field due to the presence of surface states at the superconductor's boundary. If the external electric field is sufficiently large, the resultant of the axion-mediated induced magnetic field and the Meissnerian magnetic field at $z=0$ will oppose the external magnetic field. 

Fig.~\ref{peak.fig}~(a) shows that at small tilts, the induced magnetic field is unable to overcome the external magnetic field and alter the magnetic field penetration behavior inside the superconductor.  However, as the tilt increases, even with a constant $B_0/E_0$ value, the intensity of the induced magnetic field increases, causing the trivial Meissner effect behavior to shift to the behavior of a hypergeometric function. The tilt magnitude at which this transition occurs depends on the $B_0/E_0$ ratio.

In the second panel of Fig.~\ref{peak.fig}, we plot the peak value of $B_\perp$ profile versus the tilt magnitude. This peak value directly reflects the strength of the induction field.  Notably, at point $z=0$, the induced field opposes the external magnetic field, while at the peak position they become aligned. The figure clearly demonstrates that for a fixed $B_0/E_0$ ratio, increasing the tilt amplifies both the surface state effects and the intensity of the induced magnetic field.

\section{Parallel component of the magnetic field}
\label{sec.V}
For the general case where $0 < \alpha < \pi/2$, with $\alpha$ being the angle between the applied surface electric field and the tilt direction, the electric field components take the form $E_\perp=(E_0\sin\alpha) ~ \text{exp}(-z/\lambda_B) $ and $E_\parallel=(E_0 \cos\alpha) ~ \text{exp}(-z/\lambda_B)$.

The perpendicular magnetic component $B_\perp$ exhibits the behavior analyzed in the previous section, while $B_\parallel$ shows qualitatively similar effects but without dependence on the tilt parameter $\zeta$. This results in an anisotropic magnetic field penetration, driven by the directional dependence of the tilt vector.

Fig.~\ref{anisotropy.fig} visualizes this anisotropic penetration profile for various angles $\alpha$. Panel (a) shows the distinct behaviors of $B_\perp$ and $B_\parallel$ for $\zeta=0.5$ and $\alpha=\pi/3$, while panel (b) demonstrates that the anisotropy persists even at $\alpha=\pi/4$. As previously discussed, this anisotropic response is exclusive to topological superconductors. Notably, conventional superconductors - including those with tilted Weyl cones - exhibit no such tilt-dependent signatures in their Meissner response. The observed anisotropy, therefore, provides clear evidence of topological superconductivity, with its magnitude directly correlated with the ratio of tilt components.

Figs.~\ref{anisotropy.fig}d, f, and g present the corresponding cases of panels a, b, and c for $\zeta=0$. This comparison clearly demonstrates that finite $\zeta$ enhances the anisotropic behavior. The marked contrast between $B_\perp$ and $B_\parallel$ in panels a ($\zeta=0.5$) and d ($\zeta=0$) confirms this effect. Notably, while no directional dependence appears at $\alpha=\pi/4$ when $\zeta=0$ (panel e), a clear anisotropy emerges for $\zeta=0.5$ (panel b).

As shown in Fig.~\ref{anisotropy.fig}~g, for $\theta=0$ (corresponding to a trivial Weyl superconductor), the perpendicular magnetic field component $B_\perp$ remains unaffected by $\zeta$, showing identical behavior for both $\zeta=0$ and $\zeta=0.9$. This demonstrates that the tilt can only be detected through $B_\perp$ when surface states are present, confirming the axion-mediated nature of this anisotropy.

The figure further shows that in trivial Weyl superconductors, the Meissner effect maintains its characteristic exponential decay profile regardless of $\zeta$. In contrast, for TWSCs (Fig.~\ref{anisotropy.fig}~h), $B_\perp$ exhibits a strong tilt-dependence with $\zeta$-sensitive penetration depth, while $B_\parallel$ remains completely unaffected by $\zeta$ variations.

Notably, while $B_\parallel$ displays hypergeometric behavior due to surface modes but no significant $\zeta$-dependence, $B_\perp$ shows increasing penetration depth with growing tilt. This distinct response arises from the combined effects of the axion field and tilt in topological systems.

\section{Universal Behavior}
\label{sec.VI}
As illustrated in Fig.~\ref{B.fig} (a) and (b), all curves corresponding to the same $B_0/E_0$ ratio but different tilt parameters $\zeta$ share a fixed point that is invariant under the following scaling procedure. Rescaling $z\rightarrow \Lambda z$, $\lambda_{B(\theta)}\rightarrow\Lambda \lambda_{B(\theta)}$, and $B_\perp/E_0 \rightarrow \Lambda^2 (B_\perp/E_0)$ with $\Lambda=\sqrt{1-\xi^2}$, the governing differential equation for $B_\perp$ maintains its form under $\zeta$ variation:
 \bea
-\frac{d^2 B_\perp}{dz^2} + \frac{1}{\lambda_B^2} B_x =- 
\frac{d}{dz} (e^{-z/\lambda_B} \frac{d\theta}{dz}),
\label{B-d-u}
\eea
leading to the universal magnetic field profile:
\bea
B_\perp &=& e^{-z/\lambda_B} \tilde B_0,
\label{B-univ}
\eea
with the amplitude function:
\bea
\tilde B_0 &=& \bigg\{ B_0 
+\arcsin[\tanh(z/\lambda_\theta)] 
\\
\nn
&-& \frac{2a}{a+2} e^{-z/\lambda_\theta} 
~_2F_1(1,\frac{a+2}{2a},\frac{3a+2}{2a},-e^{-2z/\lambda_\theta})
\bigg\}.
\eea

This universal scaling behavior suggests that \eqref{B-univ} may apply generally to all curved spaces, exhibiting scale invariance with respect to the tilt vector (Fig.~\ref{tilt.fig}~c).

Remarkably, at the critical tilt $\zeta = 1$, both characteristic lengths $\lambda_\theta$ and $\lambda_B$ vanish.  Consequently, the influence of surface states on the magnetic field penetration in this limit  becomes negligible, and the Meissner effect  is completely suppressed. In this scenario, the topological superconductor completely expels the applied magnetic field.

\section{Anomalous PHE and Anisotropic surface current}
\label{sec.VII}
Building on the derivation of magnetic field behavior from the previous section, we now address the planar Hall effect (PHE) at the boundaries of a tilted TWS. The presence of a non-zero axial term in topological superconductors induces an anomalous Hall effect, characterized by a surface current perpendicular to the applied electric field \cite{nogueira}. In our tilted structure, this PHE exhibits strong anisotropy attributable to the anisotropic nature of the tilt vector.

\begin{figure}[t]
\includegraphics[width=7cm]{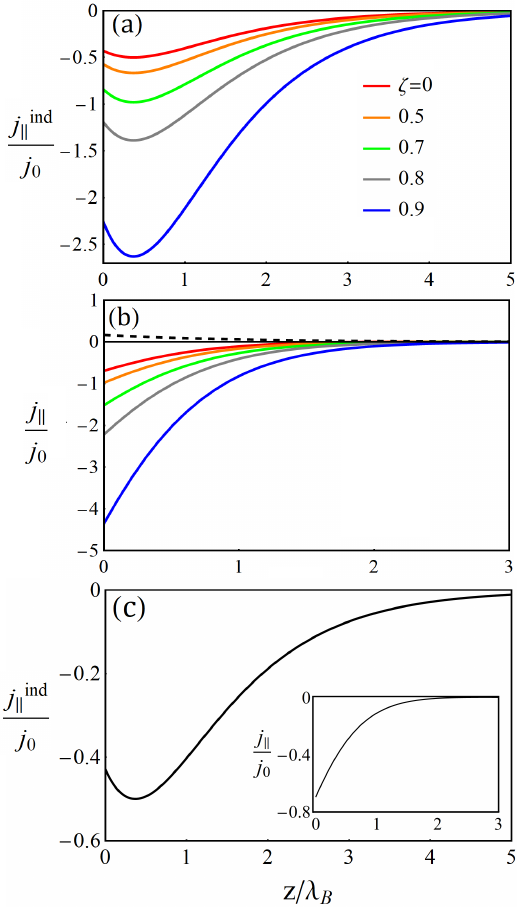}
\caption{Parallel component of the planar Hall current for $B_0=1/6~ B_{E_0}$, with $j_0=e^2 E_0 /\pi \lambda_B$. (a) The anomalous planar Hall current, showing only the electrically induced surface contribution. (b) The total surface current, including both the trivial component and the electrically induced planar Hall current in a topological superconductor.
The dashed curve represents the tilt-independent surface current in a trivial superconductor. (c) Universal scaling behavior: All curves in (a) and (b) fall into a single trend under the transformations $z\rightarrow \Lambda z$ and $\lambda_{B(\theta)}\rightarrow\Lambda  \lambda_{B(\theta)}$, with $\Lambda=\sqrt{1-\zeta^2}$. }
\label{j.fig}
\end{figure}

The current density follows from the invariant curl of $j^\nu = -2 q\Delta_0 A^\nu$ (see Appendix B for derivation), reducing to the London equation in our simplified model:
\bea
\label{j-d.eqn}
\bb\nabla\times\bb j = - \frac{1}{\lambda_B^2} \bb B,
\eea
which decomposes into perpendicular and parallel components:
\bea
&&  \frac{d^2 j_\parallel}{dz^2} - \frac{j_\parallel}{\lambda_B^2} =  
-\frac{e^2 E_\perp}{\pi (1-\zeta^2) \lambda_B^2} (\frac{d\theta}{dz})\\
\nn
&& \frac{d^2 j_\perp}{dz^2} - \frac{j_\perp}{\lambda_B^2}=\frac{e^2 E_\parallel}{\pi\lambda_B^2} (\frac{d\theta}{dz}).
\label{j_p.eqn}
\eea
Notably, $j_\parallel$ exhibits a striking distinction between trivial versus topological superconductors, mirroring the behavior of $B_\perp$. In trivial superconductors,  it shows no tilt dependence while in topological systems, it acquires $\zeta$-dependence through the axion-mediated response:
\bea
j_\parallel= 
\frac{B_\perp}{\lambda_B} 
- 
\frac{e^2 E_\perp}{\pi \lambda_B (1-\zeta^2)}
\big[\theta(z/\lambda_\theta)-\theta(0)\big],
\eea
The surface current comprises two components: a trivial component arising from the exponential decay of the applied magnetic field in the superconductor (black dashed line in Fig.~\ref{j.fig}~b), and an electrically induced planar Hall effect ($j_\parallel^{\text{ind}}$) (Fig.~\ref{j.fig}~a), unique to topological superconductors due to their surface states and absent in non-topological systems:
\bea
j_\parallel^{\text{ind}}&=& \frac{e^2 E_0 ~ e^{-z/\lambda_B}}{\pi \lambda_B (1-\zeta^2)}
\bigg\{ \arcsin{[\tanh{(z/\lambda_\theta)}]} 
\\
\nn
&& - \frac{2a}{a+2} e^{-z/\lambda_\theta} 
~_2F_1(1,\frac{a+1}{2a},\frac{3a+2}{2a},-e^{-2z/\lambda_\theta})
\\
\nn
&& - 
\big[\theta(z/\lambda_\theta)-\theta(0)\big] 
\bigg\}.
\eea

To ensure observable surface state effects in Fig.~\ref{j.fig}, we impose the condition $B_0\lesssim B_{E_0}$, adopting $B_{E_0}=B_0\approx 1 ~\text{mT}$ as a representative field strength. Given the characteristic penetration depth in Weyl superconductors of $\lambda_B \sim 100-1000 ~\text{nm}$ \cite{PhysRevB.100.220504,JAP.129.113903}, this configuration produces a characteristic current density scale of $j_0\sim 10^4-10^5 ~\text{A}/\text{m}^2$. Consequently, the perpendicular planar Hall currents fall within the range $10^3-10^4 ~\text{A}/\text{m}^2$. 

The perpendicular component of the surface current is independent of the tilt magnitude $\zeta$. Although this component includes both the trivial contribution and the electrically induced term, neither of them exhibits any tilt dependence. As a result, the perpendicular current remains entirely unaffected by the tilt.

Given that the parallel current density ($j_\parallel$) is approximately an order of magnitude larger than the  perpendicular component ($j_\perp$), a significant anisotropy emerges in the surface current. This anisotropy is solely attributed to the axion-mediated tilt effect.

In its universal form, rescaling $z\rightarrow \Lambda z$, $\lambda_B \rightarrow \Lambda \lambda_B$, and $j_\parallel \rightarrow \Lambda j_\parallel$, the governing differential equation for $j_\parallel$ maintains its form under $\zeta$ variation:
 \bea
-\frac{d^2 j_\parallel}{dz^2} + \frac{1}{\lambda_B^2} j_\parallel = 
- \bigg(\frac{e^2 E_\perp}{\pi \lambda^2_B}\bigg) 
\frac{d\theta}{dz},
\label{j-d-u}
\eea
leading to the universal form for the electrically induced current as:
\bea
j_\parallel= \frac{1}{\lambda_B} 
\bigg\{
\tilde B_0 
- 
\frac{e^2 E_0 }{\pi}
\big[\theta(z/\lambda_\theta)-\theta(0)\big]
\bigg\}
e^{-z/\lambda_B},
\eea
where the tilt dependence is incorporated through $\tilde B_0$, $\lambda_B$, $\lambda_\theta$ and $z$, with the overall scaling $j_\parallel \rightarrow j_\parallel\sqrt{1-\zeta^2}$.

Fig.~\ref{j.fig}~c shows the universal behavior of the surface current under the scaling transformations.

\section{Tilt-Dependent Modulation of superconducting gap}
\label{sec.VIII}
In our previous analysis, we considered a simplified scenario with identical, tilt-independent gap amplitudes $\Delta_0$ on both Fermi surfaces. We now generalize this treatment by introducing a tilt-dependent gap function . This modification leads to a corresponding variation in the penetration depth:
\bea
\frac{\lambda_B(\zeta)}{\lambda_B}=\frac{\Delta_0}{\Delta_0(\zeta)}
\eea
where $\lambda_B$ denotes the zero-tilt penetration depth. 

The fundamental structure of the equations remains unchanged but now incorporates $\zeta$-dependent penetration depths for both electric and magnetic fields. When expressed in terms of the scaled coordinate $z/\lambda_B(\zeta)$, all relationships retain their original form. Crucially, comparison with the tilt-independent case shows that while the functional forms are preserved, the $\zeta$-dependence quantitatively modifies the field penetration profiles—as we analyze below.

\begin{figure}[t]
\includegraphics[width=7.5cm]{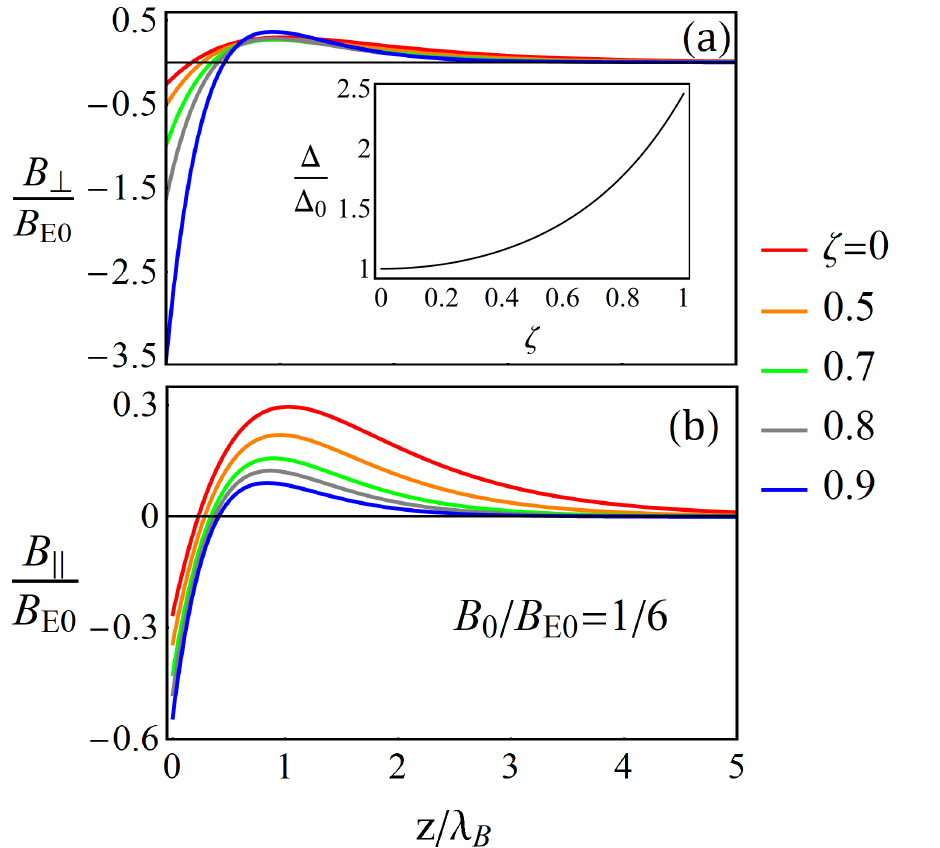}
\caption{Magnetic field penetration profiles in the TWSC including gap anisotropy effects: (a) perpendicular, $B_\perp$ and (b) parallel, $B_\parallel$ components. The horizontal axis shows the normalized distance from the surface $z/\lambda_B(0)$. Inset: Variation of the superconducting gap amplitude $\Delta_0$ with tilt parameter $\zeta$. }
\label{amplitude.fig}
\end{figure}

The system maintains s-wave superconducting pairing symmetry, which by definition preserves rotational invariance in the order parameter amplitude. As a result, the tilt parameter $\zeta$ cannot induce directional dependence in the magnitude of the superconducting gap $\Delta_0$. Adopting the exponential gap dependence $\Delta_0(\zeta)/\Delta_0$ from Ref.~\cite{tiltedWSC_Alidoust}, we calculate the magnetic response for various tilt values (Fig.~\ref{amplitude.fig}). The inset confirms our gap variation agrees with type-I Weyl superconductor (i.e., with moderate tilt) predictions.

Our calculations reveal two essential trends in the tilt-dependent response: the penetration depth of $B_\perp$ systematically decreases with increasing $\zeta$ while preserving qualitative agreement with previous results, whereas $B_\parallel$ exhibits a concurrent reduction in both its maximum field value and penetration depth as $\zeta$ increases.

By incorporating gap modulation effects, this analysis reveals significantly enhanced anisotropy in the system. Accounting for the tilt-dependent gap function $\Delta_0(\zeta)$, , we observe that $B_\perp$ exhibits progressive enhancement while $B_\parallel$ decays with increasing $\zeta$, with the growth rate of $B_\perp$ substantially exceeding the decay rate of $B_\parallel$. Both components simultaneously develop a previously unreported reduction in penetration depth. Together, these effects conclusively demonstrate how gap modulation acts to amplify the intrinsic anisotropy of the topological superconducting state.

The initial zero-tilt degeneracy between field components breaks progressively with increasing $\zeta$, with $B_\perp$ developing greater relative magnitude. This emergent anisotropy stems directly from the interplay between tilt effects and axion field coupling, as established in our earlier theoretical framework.

\section{Conclusion}
\label{sec.IX}
This work has systematically elucidated the profound influence of Weyl cone tilt on axion-mediated electromagnetic responses in topological Weyl superconductors. Through theoretical analysis of the 4D TI model, we have established that the tilt parameter $\zeta$ serves as a powerful tuning knob for the surface state electrodynamics, simultaneously modifying both the penetration depth of magnetic fields and the magnitude of the electromagnetic induction effect mediated by topological surface states. The tilt-induced enhancement of surface state contributions manifests distinctly in both the magnetic response and planar Hall currents.

Most remarkably, our analysis reveals a striking anisotropic Meissner response that emerges at finite tilt. As $\zeta \rightarrow 1$, the perpendicular magnetic field component becomes strongly dominant, while the parallel component remains finite but significantly suppressed—a clear departure from conventional superconducting behavior. This anisotropy arises exclusively from the interplay between tilt effects and axion field coupling in topological systems; no comparable anisotropy is observed in trivial Weyl superconductors. Crucially, the tilt selectively modifies the perpendicular component's penetration profile, while leaving the parallel component largely unaffected.

Equally significant is our identification of tilt-induced anisotropy in the surface Hall currents, which provides both a measurable experimental signature and a potential tool for characterizing tilt parameters in real materials. These findings collectively establish $\zeta$ as a crucial degree of freedom in engineering electromagnetic responses.

Beyond their immediate theoretical implications, these results open new possibilities for controlling superconducting electrodynamics through band structure engineering. The demonstrated sensitivity to Weyl cone tilt suggests promising avenues for developing tunable superconducting devices with anisotropic responses. Future work exploring material-specific implementations and experimental verification of these effects would be particularly valuable. We note that various other surface phenomena in real superconductors could influence the Meissner screening; a direct comparison with the experiment would require extending our model to include such effects.

\begin{appendices}

\section*{A: Reviewing the Axion field theory}
\label{app.A}
\subsection*{Mapping 3+1D TSC to 4+1D TI with s-wave Superconductor Boundaries}
  
\begin{figure}[t]
\includegraphics[width=4cm]{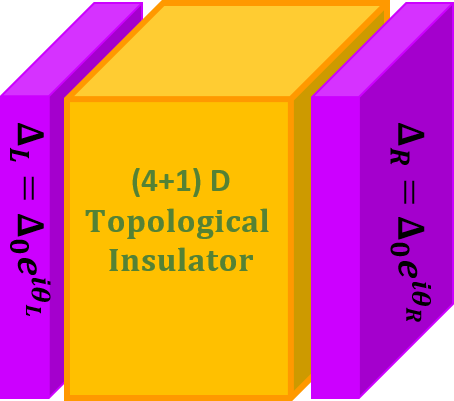}
\caption{Schematic representation of a 4+1D TI, depicted as a slab with two 3+1D boundaries on opposite sides of an interval. The boundaries are spatially separated along the extra dimension. The phase difference between the s-wave superconductors on these boundaries is mapped to a vector potential in the extra spatial dimension of the 4+1D TI. }
\label{appx.fig}
\end{figure}
To establish the mapping, we first consider a 4+1D time-reversal invariant TI which has an insulating bulk, and its boundaries support gapless surface states.The surface states of the 4+1D TI act as the relevant degrees of freedom for the TSC. The key topological invariant of this TI is its second Chern number, $C_2$ which is the coefficient of the Chern-Simon action\cite{Bernevig_Book}:
 \bea
 \nn
 S_{eff}=\frac{C_2}{4! \pi^2} \int d^4 x \int dt \epsilon^{\mu_1\mu_2\mu_3\mu_4\mu_5} A_{\mu_1} \partial_{\mu_2} A_{\mu_3} \partial_{\mu_4} A_{\mu_5}
 \eea
 which determines the Hall response as $j_{\mu_1}(x)=\delta S_{eff}[A]/\delta A_{\mu_1}(x)= (C_2/8\pi^2) \epsilon^{\mu_1\mu_2\mu_3\mu_4\mu_5} \partial_{\mu_2} A_{\mu_3} \partial_{\mu_4} A_{\mu_5}$.
 
 Consider a 4+1D TI in a slab geometry, where the bulk is bounded by two (3+1)-dimensional surfaces. These surfaces are coupled to conventional s-wave superconductors, each characterized by a superconducting order parameter with phases $\theta_L$ and $\theta_R$, as illustrated in Fig.\ref{appx.fig}. The electromagnetic field configuration is chosen as $A_x=-E_0 t$, $A_y=B_0 z$, and $A_3=A_4=A_t=0$, yielding $\bb B=F_{zy}=B_0 \hat{x}$ and $\bb E= - F_{xt}= E_0 \hat{x} $. The Lagrangian of the system is given by:
\bea
\nn
\mathcal{L} =
\mathcal{L}(\phi_1)+\mathcal{L}(\phi_1) + \mathcal{L}_J +\mathcal{L}_{EM} +\mathcal{L}_\theta,
\label{L4}
\eea
where $\mathcal{L}(\phi_i) =(\partial_\mu \phi_i)^\dagger (\partial^\mu \phi_i)+ m^2 \phi_i^\dagger \phi_i - (\frac{\lambda}{2}) (\phi_i^\dagger \phi_i)^2 $ describes the phenomenological free energy of the 3+1D superconductors on the boundaries. The term $\mathcal{L}_J=\phi_1^*\phi_2+\phi_2^*\phi_1$ arises from the Josephson coupling between the superconductors on opposite boundaries. The electromagnetic Lagrangian $\mathcal{L}_{EM}=-(1/4)F_{\mu\nu}F^{\mu\nu}$ accounts for the presence of electromagnetic fields, while $\mathcal{L}_\theta$ represents the axion term derived from the topological Chern-Simon's term.

Under a gauge transformation $A_\lambda \rightarrow \tilde{A}_\lambda= A_\lambda+\partial_\mu \Lambda $ with $\Lambda=-(1/2L_4)[\theta_L(L_4-x_4)+\theta_R x_4]$, the phases $\theta_L$ and $\theta_R$ are effectively set to zero, and the gauge field acquires an additional component $A_4\rightarrow (\theta_L - \theta_L)/2L_4$. Since the pairing terms are confined to the boundaries, the bulk remains a 4+1D TI coupled to the gauge field $\tilde{A}_\lambda$. The effective action retains the Chern-Simons term with $C_2=1$, expressed in terms of the physical degrees of freedom $\theta_L$, $\theta_R$, and $\tilde{A}_\lambda$:
 \bea
 \nn
 S_{eff}=\frac{1}{4! \pi^2} \int d^4 x \int dt \epsilon^{\mu_1\mu_2\mu_3\mu_4\mu_5} A_{\mu_1} \partial_{\mu_2} A_{\mu_3} \partial_{\mu_4} A_{\mu_5},
 \eea
 integrating over $x_4$ yields:
 \bea
 \nn
 S_{eff}=\frac{1}{32\pi^2} \int d^3 x \int dt \epsilon^{\mu_1\mu_2\mu_3\mu_4} \big( \frac{\theta_L-\theta_R}{2} \big) F_{\mu_1\mu_2}F_{\mu_3\mu_4}.
 \eea

The resulting Lagrangian describes the response of the fermions to superconducting phase fluctuations $\theta_L$ and $\theta_R$, as well as to the electromagnetic field:
 \bea
\nn
\mathcal{L} =
&& \sum_{j=L,R} \bigg[ 
(\partial_\mu\phi_j - i q A_\mu\phi_j)^\dagger ( \partial^\mu\phi_j - i q A^\mu \phi_j)   
\\
\nn
&& +~ m_j^2 |\phi_j|^2  - \frac{\lambda_j}{2} |\phi_j|^4 \bigg] \\
\nn
&&+~ \phi_L^*\phi_R + \phi_R^*\phi_L\\
\nn
&&- \frac{1}{4} F_{\mu\nu} F^{\mu\nu} - \frac{e^2(\theta_L -\theta_R)}{64\pi^2} ~ \epsilon^{\mu\nu\alpha\sigma} F_{\mu\nu} F_{\alpha\sigma}.
\eea

Setting $\phi_{L/R}=\Delta e^{i\theta_{L/R}}$ and neglecting amplitude fluctuations of the order parameters, the Lagrangian simplifies to:
\bea
\nn
\mathcal{L} =
&&\sum_{j=L,R} \bigg[ 
(\partial_\mu\theta_j - q A_\mu) (\partial^\mu\theta_j - q A^\mu ) \Delta^2 
\\
\nn
&& + ~ m_j^2 \Delta^2  - \frac{\lambda_j}{2} \Delta^4 \bigg] \\
\nn
&& + ~ 2\Delta^2 \cos{(\theta_L-\theta_R)}\\
\nn
&&- \frac{1}{4} F_{\mu\nu} F^{\mu\nu} - \frac{e^2(\theta_L -\theta_R)}{64\pi^2} ~ \epsilon^{\mu\nu\alpha\sigma} F_{\mu\nu} F_{\alpha\sigma}.
\eea

Assuming that $m_j$ and $\lambda_j$ are the same for two superconductors, 
\bea
\nn
\mathcal{L} &&=
\\
\nn
&&\sum_{j=L,R} \bigg[ 
(\partial_\mu\theta_j - q A_\mu) (\partial^\mu\theta_j - q A^\mu ) \Delta^2 \bigg] \\
\nn
&& + 2 m^2 \Delta^2  - \lambda \Delta^4 + 2\Delta^2 \cos{(\theta_L-\theta_R)}\\
\nn
&&- \frac{1}{4} F_{\mu\nu} F^{\mu\nu} - \frac{e^2(\theta_L -\theta_R)}{64\pi^2} ~ \epsilon^{\mu\nu\alpha\sigma} F_{\mu\nu} F_{\alpha\sigma}.
\eea

This expression corresponds to Eq.\eqref{main_L} in the main text. The phase difference $\theta = \theta_+ - \theta_-$ in the main text corresponds to $\theta_L - \theta_R$ here, with the identification $\theta_+ \equiv \theta_L$ and $\theta_- \equiv \theta_R$. This phase difference between the s-wave superconductors plays a central role, as it maps to a vector potential $A_4$ in the extra spatial dimension of the 4+1D TI. Through a gauge transformation, this phase difference is incorporated into the gauge field, linking the electromagnetic response of the 4+1D TI to the axion term in the effective action of the TSC.

The axion term, which couples the phase difference $\theta$ to the electromagnetic field $F_{\mu\nu}$, describes the magnetoelectric response of the TSC. This term is topological in nature and takes the form $\theta \bb E \cdot \bb B$. The resulting effective field theory for the TSC includes the axion coupling term, Maxwell term, Higgs term, and Josephson coupling terms.

\subsection*{Role of Josephson coupling }
 The effective Lagrangian for a 3D TSC includes a Josephson coupling term, among other components, which is crucial for understanding its topological properties and behavior.
In summary, the Josephson coupling is not merely a secondary interaction; it is a pivotal factor in defining the topological nature of a TSC. The sign of the Josephson coupling dictates whether the system is in a topologically non-trivial or trivial state and affects the behavior of vortices within the material. The Josephson coupling also generates an interaction between the two complex field components. The interplay between the Josephson coupling and other aspects, such as the axion term, determines the overall topological properties of the TSC.

\section*{B: contravariant vectors in curved space-time}
\label{app.B}
Consider a space-time described by the metric tensor
\bea
\nn
g_{\mu\nu}=
\begin{pmatrix}
    1-\zeta^2 & \zeta_x & \zeta_y & \zeta_z \\
    \zeta_x   &  -1     &  0     & 0 \\
    \zeta_y   &  0     &  -1     & 0 \\
    \zeta_z        &  0     &  0     & -1
\end{pmatrix},
\label{metric}
\eea
where $\zeta^2=\zeta_x^2+\zeta_y^2+\zeta_z^2$. The inverse metric $g^{\mu\nu}$ (given in Eq.~\ref{inversemetric} \cite{SaharPolariton}) allows us to raise indices and express contravariant derivatives. For instance, the time and spatial components of the gradient transform as
\bea
\nn
\partial^0=g^{0\nu}\partial_\nu
= \partial_t + \zeta_x\partial_x + \zeta_y\partial_y + \zeta_z\partial_z
\eea
and,
\bea
\nn
\partial^x 
&=& g^{1\nu}\partial_\nu\\
\nn
&=&\zeta_x \partial_t + (-1+\zeta_x^2)\partial_x +\zeta_x\zeta_y \partial_y +\zeta_x \zeta_z \partial_z.
\eea
More generally, if $\overline{\bb\nabla}$ denotes the contravariant gradient in the tilted space and $\bb\nabla$ the gradient, the relationship between them is
\bea
\nn
\overline{\bb\nabla}=\bb\zeta\partial_t - \bb\nabla -\bb\zeta(\bb\zeta\cdot\bb\nabla).
\eea
Similarly, the contravariant electromagnetic potential $\overline{\bb A}$  in the tilted frame relates to the covariant potential $\bb A$ as
\bea
\nn
\overline{\bb A}=\bb\zeta A_0 + \bb A -\bb\zeta(\bb\zeta\cdot\bb A).
\eea

For Eq.\eqref{E-u}, we note that from Eq.\eqref{B.eqn} in our model, $\bb\nabla\times(\bb\zeta\times\bb E)$ is zero for both $\bb E_\perp$ and $\bb E_\parallel$:
\bea
\nn
-(1-\zeta^2)\bb\nabla\times\bb B 
&-&\bb\nabla\times[\bb\zeta(\bb\zeta\cdot\bb B)]  
\\
\nn
&-&
\frac{1}{\lambda_B^2} [\bb A-\bb\zeta(\bb\zeta\cdot \bb A)]+\partial_t\bb E + \bb\zeta\times\partial_t \bb B\\
\nn
&=&
\frac{e^2}{8\pi} (\bb\nabla\theta \times \bb E).
\eea
we use $\bb B= \bb\nabla\times\bb A$, $ \bb\nabla\times\bb\nabla\times\bb=-\nabla^2+\bb\nabla\bb\nabla\cdot$ and $\bb E=\partial_t \bb A$, and also this fact that $\bb\nabla\cdot\bb A=0$ and in the low frequency regime where $\partial_t^2\sim\omega^2\sim 0$. Differentiating versus $t$ and using the above considerations yields:
\bea
\nn
-(1-\zeta^2) \nabla^2 \bb E 
&+&\bb\nabla\times\{ \bb\zeta[\bb\zeta\cdot(\bb\nabla\times\bb E)]\}
\\
\nn
&+&
\frac{1}{\lambda_B^2} [\bb E-\bb\zeta(\bb\zeta\cdot \bb E)]\\
\nn
&=&
0.
\eea
For the perpendicular component of $\bb E$, $\bb\zeta\cdot\bb E_\perp=0$ and $\bb\nabla\times\{ \bb\zeta[\bb\zeta\cdot(\bb\nabla\times\bb E_\perp)]\}= -\zeta^2 \nabla^2 \bb E_\perp$. So, $-\nabla^2\bb E_\perp+(1/\lambda^2)\bb E_\perp=0$.

For the parallel component $\bb E_\parallel$, $\bb\zeta(\bb\zeta\cdot\bb E_\parallel)=\zeta^2 \bb E_\parallel$  but  $\bb\zeta\cdot(\bb\nabla\times\bb E_\parallel)=0$, So $-(1-\zeta^2) \nabla^2 \bb E + (1/\lambda^2)(1-\zeta^2)\bb E=0$ which is the same as the equation for $\bb E_\perp$ for $\zeta\ne 1$. Hence we result in Eq.\eqref{E-u}.

For obtaining Eq.\eqref{B-u} we multiply $\bb\nabla\times$ from Eq.\eqref{B.eqn}. Neglecting the term $\bb\nabla\times(\bb\zeta\times\bb E)$ which is zero in our model:
\bea
\nn
(1-\zeta^2)\nabla^2\bb B 
&-&\bb\nabla\times\bb\nabla\times[\bb\zeta(\bb\zeta\cdot\bb B)]  
\\
\nn
&-&
\frac{1}{\lambda_B^2} \{ [\bb B-\bb\nabla\times[\bb\zeta(\bb\zeta\cdot \bb A)] \} \\
\nn
&=&
\frac{e^2}{8\pi} \bb\nabla\times(\bb\nabla\theta \times \bb E).
\eea
which we ignore $\bb\zeta\times\partial_t (\bb\nabla\times\bb B)\sim 0$ because  $\bb\nabla\times\bb B=\partial_t \bb E$ and $\partial_t^2 \bb E\sim 0$.

For the term $\bb\nabla\times[\bb\zeta(\bb\zeta\cdot\bb B)]$, we evaluate stepwise:
\bea
\nn
\bb\nabla\times[\bb\zeta(\bb\zeta\cdot\bb B)] &=&
\bb\nabla(\bb\zeta\cdot\bb B)\times\bb\zeta\\
\nn
&=& [(\bb\zeta.\bb\nabla)\bb B+\bb\zeta\times(\bb\nabla\times\bb B)]\times\bb\zeta\\
\nn
&=&[\bb\zeta\times(\bb\nabla\times\bb B)]\times\bb\zeta,
\eea
where we used the vector identity for $\bb\nabla(\bb a\cdot \bb b)$ and the fact that $\bb\zeta\cdot\bb\nabla=0$.The double curl operation required for Eq.\eqref{B-u} yields:
\bea
\nn
\bb\nabla\times\bb\nabla\times[\bb\zeta(\bb\zeta\cdot\bb B)] &=&
\bb\nabla\times\{[\bb\zeta\times(\bb\nabla\times\bb B)]\times\bb\zeta\}\\
\nn
&=&
-\bb\zeta\{\bb\nabla\cdot[\bb\zeta\times(\bb\nabla\times\bb B)]\}\\
\nn
&=&
\bb\zeta\{[\bb\nabla\times(\bb\nabla\times\bb B)]\cdot\bb\zeta\}\\
\nn
&=&
-\bb\zeta[(\nabla^2\bb B)\cdot\bb\zeta]\\
\nn
&=&
-\zeta^2 ~ \nabla^2\bb B_\parallel,
\eea
where we applied the vector identity $\bb\nabla\times(\bb a \times\bb\zeta)=-\bb\zeta(\bb\nabla\cdot\bb a)$, when $\bb\zeta$ is constant, and used $\bb\nabla\times(\bb\nabla\times\bb B)=-\nabla^2\bb B$. Here, $\nabla^2\bb B$ decomposed to $\nabla^2\bb B_\parallel$ (parallel to $\bb\zeta$) and $\nabla^2\bb B_\perp$ (transverse to $\bb\zeta$).

For $\bb\nabla\times[\bb\zeta(\bb\zeta\cdot \bb A)]$, we use the identity $\bb\nabla\times(\psi\bb a)=\bb\nabla\psi\times\bb a$ if $\bb a$ is a constant, so:
\bea
\nn
\bb\nabla\times[\bb\zeta(\bb\zeta\cdot \bb A)]
&=&
\bb\nabla(\bb\zeta\cdot \bb A)\times\bb\zeta\\
\nn
&=&
\bb\zeta\times(\bb\nabla\times\bb A)\times\bb\zeta
\eea
which, in summary, simplifies to Eq.\eqref{B-u}.

In the low-frequency regime and in the absence of an electrostatic potential, we set $A_0=0$ and $\partial_t=0$. By incorporating $\overline{\bb A}$ and $\overline{\bb\nabla}$ into $j^\nu$ we obtain:
\bea
\nn
\overline{\bb j}=-2 q \Delta^2 \overline{\bb A}.
\eea
Taking the curl, we derive: 
\bea
\nn
&& \bigg[ \bb\nabla+\bb\zeta(\bb\zeta\cdot\bb\nabla) \bigg]
\times
\bigg[ \bb j -\bb\zeta(\bb\zeta\cdot\bb j) \bigg]= \\
\nn
&& -\frac{1}{\lambda_B^2}  
\bigg[ \bb\nabla+\bb\zeta(\bb\zeta\cdot\bb\nabla) \bigg]
\times
\bigg[ \bb A -\bb\zeta(\bb\zeta\cdot\bb A) \bigg]
\eea
Considering the simplified model with $\bb\zeta \perp \hat z$, we have:
\bea
\nn
\bb\nabla\times\bb j 
-
\bb\zeta\times[ (\bb\nabla\times\bb j)\times\bb\zeta ]
=
-\frac{1}{\lambda_B^2}
[\bb B -\bb\zeta\times (\bb B \times\bb\zeta )]
\eea
which we use the relation $\bb\nabla\times\bb A=\bb B$. Decomposing $\bb j= \bb j_\perp + \bb j_\parallel$ and also $\bb B= \bb B_\perp + \bb B_\parallel$, and considering the fact that $\bb\nabla=\hat k \partial_z$, gives us:
\bea
\nn
&&(1-\zeta^2) \partial_z j_\parallel
=
-\frac{1}{\lambda_B^2} (1-\zeta^2) B_\perp \\
\nn
&& \partial_z j_\perp 
=
\frac{1}{\lambda_B^2} B_\parallel
\eea
which is the same as Eq.\eqref{j-d.eqn}. For Eq.~\eqref{j_p.eqn} we  get a curl from Eq.~\eqref{j-d.eqn} yielding:
\bea
\nn
\bb\nabla\times\bb\nabla\bb\times j=-\frac{1}{\lambda_B^2} \bb\nabla\times\bb B.
\eea
In our geometry, using Eq.~\eqref{B.eqn},
\bea
\nn
\bb\nabla\times\bb B =
&-& \frac{1}{1-\zeta^2}\bigg\{\bb\nabla\times[\bb\zeta(\bb\zeta\cdot\bb B)] 
\\
\nn
&+&
\frac{1}{\lambda_B^2} [\bb A-\bb\zeta(\bb\zeta\cdot \bb A)]
- \frac{e^2}{\pi} (\bb\nabla\theta \times \bb E)\bigg\} \\
\nn
=
&-& \frac{1}{1-\zeta^2}\bigg\{\bb\nabla\times[\bb\zeta(\bb\zeta\cdot\bb B)] 
\\
\nn
&+&
 \bb j-\bb\zeta(\bb\zeta\cdot \bb j)
- \frac{e^2}{\pi} (\bb\nabla\theta \times \bb E)\bigg\}
.
\eea
These equations yield, 
\bea
\nn
&&  \frac{d^2 j_\parallel}{dz^2} - \frac{j_\parallel}{\lambda_B^2} =  
-\frac{e^2 E_\perp}{\pi (1-\zeta^2) \lambda_B^2} (\frac{d\theta}{dz})\\
\nn
&& \frac{d^2 j_\perp}{dz^2} - \frac{j_\perp}{\lambda_B^2}=\frac{e^2 E_\parallel}{\pi\lambda_B^2} (\frac{d\theta}{dz}).
\eea
which is Eq.~\eqref{j_p.eqn}.

\section*{B: Non-zero $\zeta_z$ }
 \label{app.B}
 In the main text, we restricted our analysis to in-plane tilt configurations to highlight features essential for comparing topological and trivial superconducting phases. However, the inclusion of an out-of-plane component $\zeta_z$ introduces qualitatively new electromagnetic phenomena.

For the case  $\bb\zeta=(0,0,\zeta_z)$ with $\bb E_0$ and $\bb B_0$ in $x$-direction, the term $\bb\nabla\times(\bb\zeta\times\bb E)$ in Eq.\eqref{B.eqn} becomes non-negligible and generates a transverse magnetic field independent of the axion coupling. Within our established theoretical framework, Eq.\eqref{B.eqn} decomposes into:
 \bea
 \nn
 (1-\zeta_z^2)(\bb\nabla\times\bb B)_x &=& -\zeta_z^2 \partial_z B_y + \zeta_z\partial_z E_x
 \\
 \nn
 (1-\zeta^2)(\bb\nabla\times\bb B)_y &=& \zeta_z^2\partial_z B_x - \frac{1}{\lambda_B^2} A + \frac{e^2}{\pi} E_x \partial_z\theta.
 \eea
Notably, while the y-component (second line) reproduces the axion-mediated induction of Eq.\eqref{B-u}, the x-component (first line) reveals a distinct tilt-induced mechanism:
\bea
\nn
B_y = - ~ \frac{\zeta_z E_0}{(1-2\zeta_z^2)\lambda_B^2} e^{-z/\lambda_B}.
\eea
This perpendicular field response represents a universal characteristic shared by both topological and trivial superconductors, rather than being unique to systems with topological surface states.

\end{appendices}
 
\bibliography{mybib}

\end{document}